# A Local criterion of topological phase transitions

*Yangfan Hu\*,*

*Research Institute of Interdisciplinary Science & School of Materials Science and Engineering, Dongguan University of Technology, Dongguan, 523808, China.*

**Abstract**

A local criterion of topological phase transitions is established based on the Morse theory: a topological phase transition occurs when the count of Morse critical points of the order function changes. The locations in space where this change occurs are referred to as spatial critical points of the topological phase transition. In cases of continuous topological phase transitions, these spatial critical points are identified through the emergence of degenerate Morse critical points, where local maxima and minima of the order function split or merge. This resembles the formation and annihilation of a particle-antiparticle pair. The wide-ranging applicability of this criterion is demonstrated through three case studies that explore topological phase transitions in both configuration space and reciprocal space. Every topological phase transition is linked to a localized physical process that cannot be comprehended solely by studying changes in a global quantity, such as a topological invariant.

\* Corresponding author: huyf@dgut.edu.cn

Topological phase transitions[1-5] are a fascinating area of research that explores the transformation between different states of matter with distinct topological properties. Unlike traditional phases like solid, liquid, and gas, topological phases[6-12] are characterized by nontrivial topological features of the system. For ordinary phase transitions, the mechanism of symmetry breaking [13, 14] enables us to select coefficients from group representation theory as order parameters, since they are zero on one side of the transition and non-zero on the other. Similarly, during a topological phase transition, the order parameter is typically chosen as a global topological invariant of the system's state variables, being zero on one side of the topological phase transition and non-zero on the other side [15].

Topological invariants represent global characteristics of a system, whereas topological phase transitions are always linked to specific localized physical processes. For instance, consider a system containing a certain concentration of topological solitons, and assume that it undergoes a topological phase transition, which indicates a change in the topological density, or more precisely, the density of topological solitons. At the microscopic level, this inevitably entails the creation or annihilation of some topological solitons at specific spatial points within the system, which cannot be revealed in the computation of topological invariants. This raises the question of whether topological invariants are indeed a sufficient tool for describing topological phase transitions.

The occurrence of ordinary phase transitions, on the other hand, rely on the stability of two or more minima in the system's free energy functional and their relative positions in terms of the value of free energy. Consequently, the determination of critical conditions for these phase transitions is predominantly determined by the second-order variation of the free energy functional in terms of the order parameters at these minimum states [16]. Applying this condition necessitates that the free energy is completely described by the order parameters [13, 14]. However, topological invariants appear only in specific terms of the free energy (e.g., the Wess-Zumino term in a 2D nonlinear sigma model[15]). This complicates the determination of whether a topological phase transition occurs based on ordinary second-order variations. Indeed, we are only able to retrospectively calculate the topological invariant of the initial and final states for any given physical process to ascertain the presence of a topological phase transition. This raises the question of whether topological phase transitions align with an autonomous localized criterion and how this criterion harmonizes with the second-order variation analysis of ordinary phase transitions.

By definition, two fields are considered topologically equivalent if they can be mapped to each other through a "continuous deformation". Consider a change of 1D scalar field from $y_1(x)$ to $y_2(x)$, if this change is topologically protected, we have $y_2(x) = y_1(f(x)) = y_1(x - u(x))$, where $u(x)$ is called an emergent displacement field (see Supplementary Materials for a detailed discussion), which yields

$$f(x) = y_1^{-1}(y_2(x)), \tag{1}$$

or

$$u(x) = x - y_1^{-1}(y_2(x)). \tag{2}$$

If for every $x$, $f(x)$ in eq. (1) satisfies $f'(x) \in (0, \infty)$, or $u(x)$ in eq. (2) satisfies $u'(x) \in (-\infty, 1)$, the change $y_1(x) \to y_2(x)$ is topologically protected according to continuum mechanics[17].

Eqs. (1, 2) are not easy to apply because it is difficult to analytically derive the inverse function of an arbitrary function. A more practical situation is when we can find a continuous variation path between $y_1(x)$ and $y_2(x)$, or in other words there exists a group of functions $g(x, b)$ which varies continuously with the control parameter $b$, such that $g(x, b_1) = y_1(x)$ and $g(x, b_2) = y_2(x)$. Hereafter, we call $g(x, b)$ the order function. The group of functions $g(x, b)$ between $b = b_1$ and $b = b_2$ describes the actual physical process of the change $y_1(x) \to y_2(x)$. Assume that $g(x, b_l)$ is topologically equivalent to $g(x, b_r)$ for any two $b_l, b_r \in [b_1, b_2]$, we then have for any $b \in [b_1, b_2]$: $\lim_{\Delta b \to 0} g(x, b + \Delta b) = g(x - \delta u_b(x), b) = g(x, b) - g'(x, b)\delta u_b(x)$, where $\delta u_b(x)$ denotes the infinitesimal emergent displacement field that links $\lim_{\Delta b \to 0} g(x, b + \Delta b)$ and $g(x, b)$, and $g'(x, b) = \frac{dg(x,b)}{dx}$. We then have after manipulation

$$\delta u_b'(x) = -\frac{\delta g'(x, b)}{g'(x, b)} + \frac{\delta g(x, b) g''(x, b)}{g'(x, b)^2}, \tag{3}$$

where $\delta g(x, b) = \lim_{\Delta b \to 0} g(x, b + \Delta b) - g(x, b)$. Eq. (3) shows that at those points where $g'(x, b) = 0$, the requirement $\delta u_b'(x) \in (-\infty, 1)$ cannot be satisfied. For a $n$-dimensional scalar function $g(\mathbf{x}, b)$, this condition is (see SM for derivation)

$$\nabla g(\mathbf{x}, b) = \mathbf{0}. \tag{4}$$

In the Morse theory[18, 19], points satisfying eq. (4) are called Morse critical points (MCPs) of $g(\mathbf{x}, b)$, in order to be distinguished from the critical point of a phase transition. The above analysis provides a variational explanation for the significance of MCPs when concerning the topological structure of a function. Taking a 1D scalar function as an example, any function can be divided into a series of monotonic intervals based on its MCPs. Two functions are considered topologically equivalent if they share the same count of MCPs and the same sequence of their appearance. Consequently, the monotonic intervals that correspond to these two functions can be smoothly and sequentially mapped onto each other. In this sense, the order function $g(\mathbf{x}, b)$ that changes continuously with the control parameter $b$ is a topologically protected set if the number of MCPs is conserved for $b$. And a local criterion of continuous topological phase transition can be proposed accordingly: <u>a critical value of control parameter $b_c$ for a continuous topological phase transition of $g(\mathbf{x}, b)$ is reached if the number of MCPs differs for $g(\mathbf{x}, b_c - \varepsilon)$ and $g(\mathbf{x}, b_c + \varepsilon)$ as $\varepsilon \to 0$. At the critical value $b = b_c$, there exist at least one spatial critical point $\mathbf{x} = \mathbf{x}_c$ at which a degenerate MCP appears such that</u>

$$\left|\nabla^2 g(\mathbf{x}, b_c)\right|_{\mathbf{x}=\mathbf{x}_c} = 0, \tag{5}$$

where $\nabla^2 g(\mathbf{x}, b_c)$ denotes the Hessian matrix of $g(\mathbf{x}, b_c)$ with components $\frac{\partial^2 g(\mathbf{x}, b_c)}{\partial x_i \partial x_j}$. To develop an intuitive grasp of the localized essence of topological phase transitions, two instances are showcased where the count of MCPs shifts at a spatial critical point within a continuous topological phase transition of a 1D scalar function:

a. Before the transition, the spatial critical point is not a MCP. Post-transition, two MCPs (a pair of local minimum and local maximum) emerge from it (Fig. 1(a)).

b. Before the transition, the spatial critical point is an MCP. Post-transition, three MCPs (including an addition of a pair of local minimum and local maximum) emerge from it (Fig. 1(b)).

For topological phase transitions that are not continuous, the above local criterion still holds but eq. (5) may not be valid.

*Consistency between the local and global criterion for topological phase transitions.*

The Euler characteristic $\chi_b$, a topological invariant of the function $g(\mathbf{x}, b)$ can be expressed as[18, 19]

$$\chi_b = \sum_{i=1}^{m} (-1)^i \mu_i(b), \tag{6}$$

where $\mu_i(b)$ ($i = 1, 2, ..., n$) denotes the Morse numbers with index $i$, i.e., the numbers of MCPs of $g(\mathbf{x}, b)$ with $i$ negative eigenvalues of the Hessian matrix $\nabla^2 g(\mathbf{x}, b_c)$. Eq. (6) establishes the connection between alterations in the MCPs count and the global topological invariant of the function. Consequently, changes in the MCPs count inherently lead to sudden shifts in the topological invariant, guaranteeing the consistency between our proposed criterion for local topological phase transitions and the global criterion. Simultaneously, the criterion for local topological phase transitions clarify that shifts in the topological invariant arise from variations in the MCPs count. The application of Eq. (6) requires the order function $g(\mathbf{x}, b)$ to be a Morse function, which is not always the case. Nonetheless, we can always transform it into a Morse function through infinitesimal perturbations[20], making this a negligible limitation.

*Comparison between the second-order variation criterion for ordinary phase transitions and the local criterion for topological phase transitions.*

From the above analysis, we discern that topological phase transitions offer a criterion that parallels ordinary phase transitions. The benchmark for the latter corresponds to a vanishing second-order variation of the free energy, indicating instability in solutions, while the benchmark for the former corresponds to alterations in the count of solutions' MCPs, signifying changes in the solutions' topological structure. Within this framework, phase transitions can be categorized into three types: a. those that are both ordinary and topological phase transitions (solutions become unstable and evolve into new solutions with changed topological structure); b. those that are only topological phase transitions (solutions remain stable but experience a change in topological structure); c. those that are only ordinary phase transitions (solutions become unstable and evolve into new solutions without altering the topological structure). In a subsequent study on phase transitions in chiral magnets [21], we present specific examples of these three types of transitions. Another thought-provoking subject is why the overall properties of a system change during a topological phase transition even when solutions remain stable. Drawing inspiration from the two examples illustrated in Fig. 1, the disparity on either side of a topological phase transition stems from the emergence of 1-2 new monotonic intervals in the corresponding function. This, in turn, leads to variations in the overall response of the function to the control parameter.

*Analysis of topological phase transitions based on free energy density order function.*

Different physical systems are described by distinct physical quantities, but they can be compared by calculating the free energy density. Consequently, a universal approach to analyze topological phase transitions in different systems can be established by choosing the free energy density as the order function. In general, a topologically nontrivial state corresponds to an inhomogeneous spatial distribution of free energy density. Thus, a transition from a topologically trivial phase to a topologically nontrivial phase corresponds to a change of free energy density distribution from a spatially uniform one to a spatially nonuniform one.

Consider a one-dimensional system positioned within the bistability region, where two locally stable solutions of the free energy minima coexist. Solution A is characterized by a uniformly distributed free energy density, while solution B exhibits a peak and a valley in the free energy density (illustrated in Fig. 2). The system's preference for one solution over the other is determined by competition between the area of the peak region (area $P$) and the area of the valley region (area $V$) with reference to solution A's free energy density (Fig. 2). Specifically, if area $P >$ area $V$, then solution A is favored by the system; conversely, if area $P <$ area $V$, then solution B is favored. Therefore, identification of this phase transition as a topological one is accomplished by marking the new maximum and minimum points in state B compared with state A (purple triangle marks in Fig. 2), while the occurrence of this topological phase transition is attributed to the breaking of symmetry in the peak-valley area of the difference in free energy density between the two phases.

*Ordinary phase transitions within the Landau mean-field approximation as topological phase transitions in the configuration space.*

In the study of topological phase transitions, the order function $g(\mathbf{x}, b)$ is usually chosen as the state variable of interest of the system, while the spatial variable $\mathbf{x}$ denotes the real space or, when studying the collective modes in crystals, the reciprocal space. Alternatively, for spatially homogeneous systems, we can also choose $g(\mathbf{x}, b)$ to be the free energy while $\mathbf{x}$ represents the configuration space. For example, the classic Landau theory of phase transitions[13, 14] describes the free energy as:

$$F(M,T) = \alpha(T - T_0)M^2 + \beta M^4, \qquad (7)$$

where $g(x,b)$ is here the free energy $F(M,T)$, the spatial variable $x$ is the order parameter $M$, and the control parameter $b$ is the temperature $T$. As the solution $M=0$ is homogeneous in space, the critical condition for ordinary phase transitions is given by $\left(\frac{\partial^2 F}{\partial M^2}\right)_{M=0}=0$, which yields $T_c=T_0$. On the other hand, the critical condition for topological phase transitions requires a sudden change in the count of MCPs of the function $F(M,T)$ on both sides of $T=T_c$, which yields $T_c=T_0$. According to Eq. (5), at the spatial critical point, $\left(\frac{\partial^2 F(M,T_c)}{\partial M^2}\right)_{M=M_c}=0$, which determines $M_c=0$ as the spatial critical point. Thus, we realize that the ordinary phase transitions in the Landau mean-field approximation can be viewed as topological phase transitions in the configuration space. Moreover, in this scenario, the condition provided by Eq. (5) is essentially a transformation of the second-order variation condition for ordinary phase transitions: the second-order variation condition determines the critical condition for a known solution, while Eq. (5) determines the solution for a known critical condition.

*Two examples of topological phase transitions in the reciprocal space*

- Phonon spectrum of a diatomic chain

Topological phase transitions are intrinsically local, but if we replace the real space by the reciprocal space, a local point now represents one or several global dynamical modes, which enables studies of topological changes of global properties. Within the harmonic approximation, the lattice dynamics of a crystalline structure with arbitrary complexity can be reduced to a set of independent phononic branches with their own dispersion relations. For a 1D biatomic chain[22], we have

$$\begin{cases} \omega_a^2(q,b) = \beta\frac{1+b}{bm}\left(1-\sqrt{1-\frac{4b}{(1+b)^2}\sin^2 aq}\right), \\ \omega_o^2(q,b) = \beta\frac{1+b}{bm}\left(1+\sqrt{1-\frac{4b}{(1+b)^2}\sin^2 aq}\right), \end{cases} \quad (8)$$

where $\omega_a(q,b)$ denotes the dispersion relation of the acoustic branch, $\omega_o(q,b)$ denotes the dispersion relation of the optical branch, $b$ is the mass ratio between the heavy and light atom ($b \geq 1$), and $m$ is the mass of the light atom. For convenience, by setting $\beta=m=a=1$ (Fig. 3(a)), we have from eq. (8)

$$\begin{cases} \omega'_a(q,b) = \dfrac{\sin 2q}{f(q)\sqrt{1+(1-f(q))/b}}, \\ \omega'_o(q,b) = \dfrac{\sin 2q}{f(q)\sqrt{1+(1+f(q))/b}}, \end{cases} \quad (9)$$

where $f(q) = \sqrt{1+b^2+2b\cos 2q}$. We study the distribution of $\omega'_a(q,b)$ and $\omega'_o(q,b)$ in the Brillouin zone at different values of $b$ (Fig. 3(b)). Choosing both $\omega_a(q,b)$ and $\omega_o(q,b)$ as the order functions, a topological phase transition occurs at $b_c = 1$. At $b_c = 1$, a biatomic chain reduces to a monoatomic chain, where the band gap between the acoustic branch and the optical branch is closed. However, the spatial critical point cannot be obtained by using Eq. (5), because Eq. (8) corresponds to an interesting where we do have analytical expression of the order functions but they do not change continuously with the control parameter $b$. It is precisely at the critical condition $b_c = 1$ that the change in the order functions becomes discontinuous at $q = \frac{\pi}{2}$, and leads to a well known band-gap opening. According to our theory, it is the appearance of the new MCPs at both side of $q = \frac{\pi}{2}$ instead of the band gap opening itself that causes the topological phase transitions. It can be proved that $q = \frac{\pi}{2}$ serves as the spatial critical point corresponding to this topological phase transition.

- Band structure of a monoatomic chain within the Kronig-Penny model

We then study the band structure of a monoatomic chain within the Kronig-Penny model, whose band dispersion relation is determined by [23]

$$(P/Ka)\sin Ka + \cos Ka = \cos qa, \quad (10)$$

where $P$ is a constant representing the strength of periodic potential and $a$ is the lattice constant. We can choose the eigenvalue of energy $\epsilon(q,P) = \hbar^2 K(q,P)^2/2m$ as the order function and $P$ as the control parameter, where the dispersion relation $K = K_0(q,P)$ can be numerically solved from eq. (10). Setting $a = 1$, we show in Fig. 4 the dispersion relation as well as the variation of $\epsilon'(q,P)$ with $q$ for different values of $P$ in $q \in (0,4\pi)$. At $P = 0$ eq. (10) has a simple solution $K = q$ with no MCP. As $P$ increases from zero, there is band gap opening at $q = i\pi$, $(i = 0, 1, ...)$. It is shown that on both sides of the boundary of a band gap, we have $\epsilon'(q, P > 0) = 0$, which yields a topological phase transition with critical condition $P_c = 0$. Again the variation of order function at $P_c = 0$ is not continuous and we have multiple spatial critical points which are $q =$

$i\pi$, $(i = 0, 1, ...)$.

**Acknowledgments**

The work was supported by the NSFC (National Natural Science Foundation of China) through the funds with Grant Nos. 12172090, 11772360, 11832019.

**Author contributions**

Y.H. conceived the idea and conducted the work.

**Competing interests**

The authors declare no competing interests.

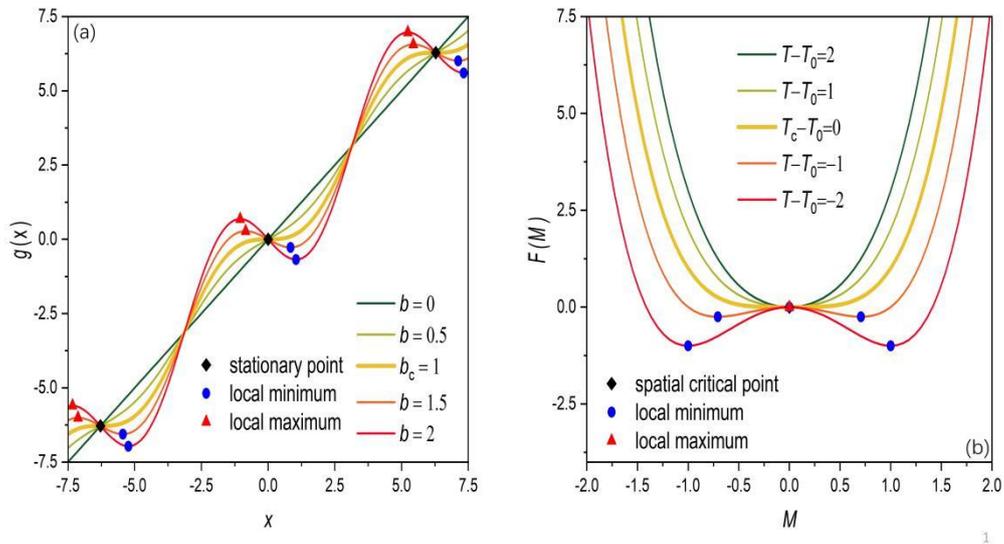

Figure 1. Image of (a) the function $g(x, b) = x - b \sin x$ for different values of $b$, and (b) the function $F(M, T) = (T - T_0)M^2 + M^4$ for different values of $T - T_0$, where $F(M, T)$ is the order function, $M$ is the spatial variable and $T$ is the control parameter.

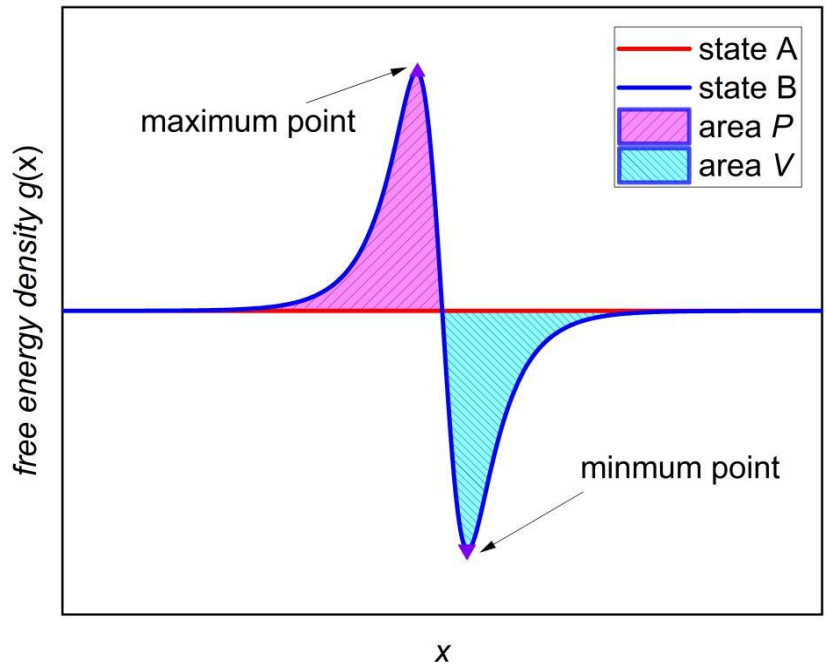

Figure 2. The free energy density distributions corresponding to two minimal states with equal free energy: (a) in state A, the free energy density is homogeneous in space, (b) in state B, there is a peak and a valley in the free energy density in space.

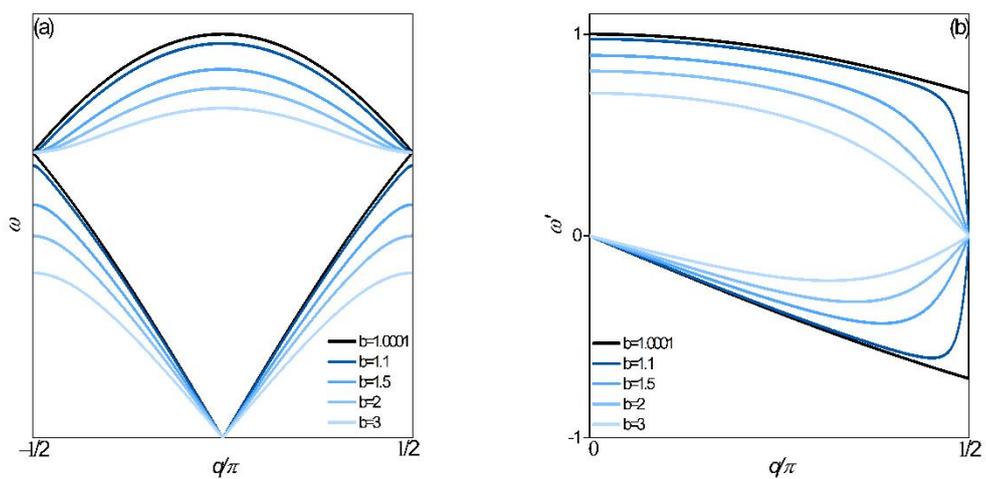

Figure 3. (a) Dispersion relation and (b) variation of $\frac{d\omega}{dq}$ with $q/\pi$ of a diatomic chain with different parameters of mass ratio $b$.

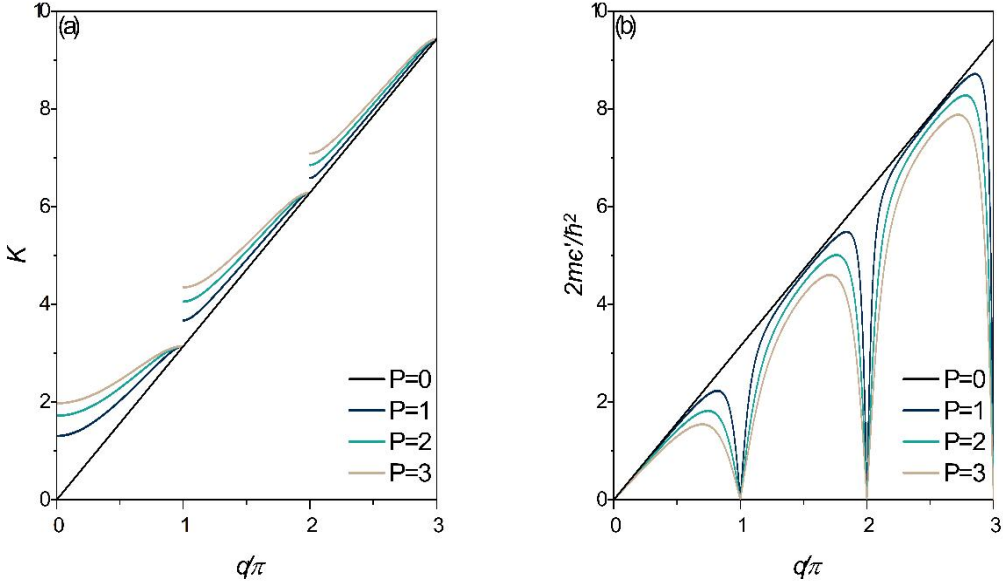

Figure 4. (a) Band structure and (b) variation of $\frac{2m}{\hbar^2}\frac{d\epsilon}{dq}$ with $q/\pi$ of a mono-atomic chain calculated at different strength of the periodic potential $P$.

**Supplementary Materials**

*A. What is an emergent displacement?*

Consider a 1D scalar field $y(x)$ undergoing a topologically protected change, so that after the change, the new field can be expressed as a "continuous deformation" of the underlying space of the old field. According to continuum mechanics[S1], the new field can be expressed by $y(f(x))$, where $f(x)$ denotes a smooth and one-to-one function. A more familiar form of expression of the new field is $y(x - u(x))$, where $u(x) = x - f(x)$ is called an emergent displacement field in the Euler coordinates, in order to be distinguished from the real displacement field studied in solid mechanics. As an example, we compare the function image of $y(x) = x, u(x) = 0$ (Figure S1(a)) and $y(x) = x, u(x) = \sin x$ (Figure S1(c)), and the effect of $u(x)$ is presented by the inhomogeneous distribution of vertical lines (geodesics) in Figure S1(c).

Similarly, we can also define the emergent displacement field $v(y)$ in the range, that is $y(x) \to (y + v(y))(x)$ (Figure S1(d)), and a more general case where both the domain and the range undergo emergent displacements, that is $y(x) \to (y + v(y))(x - u(x))$ (Figure S1(b)). For $y(x)$ with unbounded domain and range, a topologically protected change of $y(x)$ does not correspond to a unique choice of $u(x)$ and $v(y)$. For example, the function images illustrated in Figure S1(c) and Figure S1(d) are the same, but they come from different choices of $u(x)$ and $v(y)$. When studying topological stability, we mainly focus on $u(x)$, so a proper and general choice of $v(y)$ is $v(y) = a \cdot y + b$, with $a$ and $b$ constants, and the effect of $v(y)$ is to describe the change of range in a homogeneous way. For convenience, we will assume that $v(y) = 0$ in the following.

For $u(x)$ or $f(x)$ to describe a "continuous deformation", we have $u'(x) \in (-\infty, 1)$ or $f'(x) \in (0, \infty)$ for any $x$, where $u'(x) = \frac{du}{dx}$. In the example $u(x) = \sin x$, we have $u'(x) = 1$ at $x = n\pi, (n = 0, \pm 1, \pm 2, \ldots)$. It is shown in Figure S1(c) that at these points (MCPs in the Morse theory [18, 19]) the relative elongation locally reaches infinity (i.e., at these points the space is locally teared apart).

### B. Derivation of the condition of MCPs for a n-dimensional scalar field based on emergent elasticity

Consider a set of 2D scalar function $g(x_1, x_2, b)$ that change continuously with the control parameter $b$, we define
$$\delta g(x_1, x_2, b) = \lim_{\Delta b \to 0} g(x_1, x_2, b + \Delta b) - g(x_1, x_2, b). \tag{S1}$$

If the topology is protected, we have
$$\begin{aligned}\delta g(x_1, x_2, b) &= g(x_1 - \delta u_1(x_1, x_2), x_2 - \delta u_2(x_1, x_2), b) - g(x_1, x_2, b) \\ &= -\frac{\partial g}{\partial x_1} \delta u_1(x_1, x_2) - \frac{\partial g}{\partial x_2} \delta u_2(x_1, x_2).\end{aligned} \tag{S2}$$

Please note that the number of function variables on both sides of Eq. (S2) is not equal. Recall that in the discussion of Part A, a description of the variation of a function $y(x) \to (y + v(y))(x - u(x))$ does not correspond to a unique choice of $u(x)$ and $v(y)$ (e.g., the emergent deformation corresponding to the variation $y(x) = x \to x - \sin x$ could be chosen as $\begin{cases} u(x) = \sin x \\ v(y) = 0 \end{cases}$ or $\begin{cases} u(x) = 0 \\ v(y) = -\sin y \end{cases}$). The situation corresponding to Eq. (S2) is similar to this. Therefore, in all equivalent descriptions $\begin{cases} \delta u_1(x_1, x_2) \\ \delta u_2(x_1, x_2) \end{cases}$ concerning $\delta g(x_1, x_2, b)$, as long as there exists a description satisfying the definition of a continuous emergent deformation, $g(x_1, x_2, b)$ is topological protected at $b$. Based on the definition of in equation (S1), it can be understood that $\delta g(x_1, x_2, b)$ is an infinitesimal function. Therefore, if both $\frac{\partial g}{\partial x_1}$ and $\frac{\partial g}{\partial x_2}$ are non-zero, there always exists infinitesimal $\begin{cases} \delta u_1(x_1, x_2) \\ \delta u_2(x_1, x_2) \end{cases}$ such that Eq. (S2) holds, clearly indicating a continuous emergent deformation. If at a certain point $\mathbf{x} = \mathbf{x_c}$, $\frac{\partial g}{\partial x_1} = 0$ whereas $\frac{\partial g}{\partial x_2} \neq 0$, it can be set that $\delta u_1(x_1, x_2) = 0$, and $\delta u_2(x_1, x_2) = -\delta g(x_1, x_2, b)/\frac{\partial g}{\partial x_2}$, this is also a form of continuous emergent deformation. Hence, only when at a point $\mathbf{x} = \mathbf{x_c}$, $\frac{\partial g}{\partial x_1} = 0$ and $\frac{\partial g}{\partial x_2} = 0$, no description of $\begin{cases} \delta u_1(x_1, x_2) \\ \delta u_2(x_1, x_2) \end{cases}$ fulfills the definition of a continuous emergent deformation. In other words, a condition for the existence of a MCP at a point $\mathbf{x} = \mathbf{x_c}$ is that:

$$\nabla g(x_1, x_2, b) = \frac{\partial g}{\partial x_1}\mathbf{e}_1 + \frac{\partial g}{\partial x_2}\mathbf{e}_2 = \mathbf{0}. \tag{S3}$$

Clearly, the analysis of Eq. (S3) can be extended to any *n*-dimensional scalar field, which yields Eq. (4).

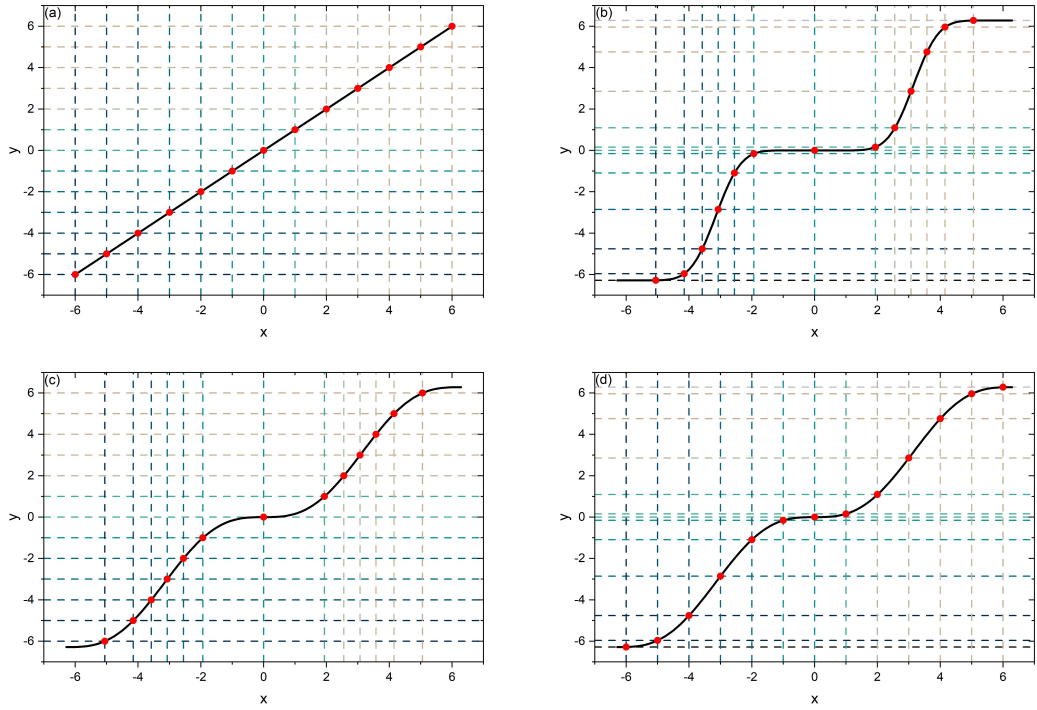

Figure S1. Image of the function $(y + v(y))(x - u(x))$, where (a) $y(x) = x, u(x) = 0, v(y) = 0$, (b) $y(x) = x, u(x) = \sin x, v(y) = \sin y$, (c) $y(x) = x, u(x) = \sin x, v(y) = 0$, and (d) $y(x) = x, u(x) = 0, v(y) = \sin y$.

[S1] *Gurtin, M. E. (1982). An introduction to continuum mechanics. Academic press.*